\documentclass[twocolumn,showpacs,superscriptaddress,prl,showkeys]{revtex4}

\usepackage{graphicx}
\usepackage{dcolumn}
\usepackage{bm}
\usepackage{amsmath,amssymb}

\def\be{\begin{equation}}
\def\ee{\end{equation}}
\def\e#1{\label{#1}\end{equation}}
\def\bea{\begin{eqnarray}}
\def\eea{\end{eqnarray}}
\def\ea#1{\label{#1}\end{eqnarray}}

\def\bem#1{\begin{mathletters}\label{#1}}
\def\eml{\end{mathletters}}

\def\ket#1{{|#1\rangle}}

\def\4#1{{\boldsymbol{#1}}}
\def\8#1{{\widetilde{#1}}}

\begin{document}

\title{Generalized Quantum State Sharing}

\author{Goren Gordon}
\affiliation{Department of Chemical Physics,
Weizmann Institute of Science, Rehovot 76100, Israel}

\author{Gustavo Rigolin}
\email{rigolin@ifi.unicamp.br}
\affiliation{Departamento de F\'{\i}sica da Mat\'eria Condensada,
Instituto de F\'{\i}sica Gleb Wataghin, Universidade Estadual de
Campinas, C.P.6165, cep 13084-971, Campinas, S\~ao Paulo, Brazil}

\date{\today}

\begin{abstract}
We present two quantum state sharing protocols where the channels are not
maximally entangled states. By properly choosing the measurement basis it
is possible to achieve unity fidelity transfer of the state if
the parties collaborate. We also show that contrary to the protocols where
we have maximally entangled channels these protocols are probabilistic. We
then compare the efficiency of both protocols and sketch the generalization
of the protocols to $N$ parties.
\end{abstract}


\maketitle

One of the most useful tools in quantum communication is the ability of a
sender (Alice) to transfer a quantum state (qubit) to a specific receiver
(Bob or Charlie) if both Bob and Charlie collaborate to recover the state
\cite{Hil99,Lo99,Pen04,Lan04,Den05a,Den05b}.
The important feature of such a scheme is that
at the end of the protocol the information contained in the transferred state
is completely available to only one of the parties and Alice is free to choose
whether Bob or Charlie will be the receiver. This controlled transmission of a
quantum state was called Quantum State Sharing (QSTS) by Lance
\textit{et al.} \cite{Lan04} to differentiate from the
controlled sharing of classical information via quantum channels, i. e.
Quantum Secret Sharing (QSS) \cite{Hil99,Kar99,Gis01}.

Many quantum information tasks require a secure transmission of quantum states.
One example, as noted in Ref. \cite{Lan04}, is quantum information networks
\cite{Cir97}, which are built of nodes in which quantum states are created,
manipulated, and stored. These nodes are connected by quantum channels and
QSTS could be employed to avoid errors and eavesdropping during the
transmission of a state between nodes \cite{Lan04,Lo99}.

All the QSTS protocols to date are based on maximally bipartite or multipartite
entangled states. In Refs. \cite{Pen04,Den05b} the quantum channels are Bell
states, $|\Psi^{-}\rangle = (1/\sqrt{2})(|01\rangle + |10\rangle)$
for example, and in Refs. \cite{Hil99,Den05a} we have
Greenberger-Horne-Zeilinger (GHZ) states, i. e.
$|GHZ\rangle  = (1/\sqrt{2})(|000\rangle + |111\rangle)$.

In a realistic situation, however, decoherence and noise degrade the channel
and we do not have a maximally entangled state anymore. One way out of this
problem is to employ quantum distillation protocols \cite{Ben96},
which allow us to obtain a maximally entangled state from a large
ensemble of partially entangled states.

Even though quantum distillation is useful to increase the entanglement of a
quantum channel it is useless if we do not have an ensemble of partially
entangled states. In addition to this,  we should note that quantum
distillation only achieves a maximally entangled state asymptotically.
Thus, for finite runs of the distillation protocol we always obtain an
almost maximally entangled state.

In view of that we are led to ask if it is possible to implement
QSTS using partially entangled states from the start. In this contribution we
show that it is indeed possible to construct such protocols. Furthermore,
the shared quantum state reaches its destination with unity fidelity. The
price we pay to achieve unity fidelity is that the protocol is no more
deterministic.

Inspired by the probabilistic quantum teleportation protocol of Agrawal and
Pati \cite{Agr02} we present two QSTS protocols. The first one uses
non-maximally entangled GHZ states as the channel and it is a generalization
of the QSTS protocol presented in Ref. \cite{Hil99}. The second one uses
non-maximally entangled Bell states as the channel and it is based on the
protocol presented in Ref. \cite{Pen04}. We then relax the requirement of
unity fidelity and employing the techniques developed in Ref. \cite{Gor06} we
compare and discuss the efficiency of both protocols.

Let us assume Alice wants to transfer to Bob or Charlie the state
$
|\phi\rangle = \alpha |0\rangle + \beta |1\rangle,
$
with $\alpha$ and $\beta$ complex and $|\alpha|^2 + |\beta|^2 = 1$. The first
probabilistic QSTS protocol can be constructed as follows. Alice shares with
Bob and Charlie the state
$
|GHZ_{n}\rangle = N (|000\rangle + n |111\rangle),
$
where $n$ can be complex and $N=1/\sqrt{1+|n|^2}$. The first qubit
belongs to Alice, the second one to Bob, and the last one to Charlie. Note
that here we allow $n$ to be any complex number and only for $n=1$ we recover
the Hillery \textit{et al.} channel. The initial state can be written as
\begin{equation}
|\Phi\rangle = | \phi \rangle_A \otimes |GHZ_n\rangle_{ABC}.
\label{initial-ghz}
\end{equation}
The subindices are written to highlight which qubit is with Alice (A), Bob (B),
and Charlie (C).
If we define the generalized Bell basis \cite{Agr02,Gor06}
\begin{eqnarray}
|\Phi^{+}_{m}\rangle &=& M (|00\rangle + m |11\rangle),\label{Bell1}\\
|\Phi^{-}_{m}\rangle &=& M (m^*|00\rangle - |11\rangle),\\
|\Psi^{+}_{m}\rangle &=& M (|01\rangle + m |10\rangle),\\
|\Psi^{-}_{m}\rangle &=& M (m^*|01\rangle - |10\rangle),\label{Bell2}
\end{eqnarray}
where $M=1/\sqrt{1 +|m|^2}$ we introduce, as will become clear soon,
a free parameter ($m$) in the protocol. It is a proper manipulation of this
parameter which makes the protocol work. Using
Eqs.~(\ref{Bell1}) to (\ref{Bell2}) we can express Eq.~(\ref{initial-ghz}) as
\begin{eqnarray}
|\Phi\rangle &=& NM \left[ |\Phi^{+}_{m}\rangle
\left( \alpha |00\rangle + m^*n\beta|11\rangle\right) \right.\nonumber \\
& & + |\Phi^{-}_{m}\rangle \left( m\alpha |00\rangle - n\beta|11\rangle\right)
 \nonumber \\
& & + |\Psi^{+}_{m}\rangle \left( n\alpha |11\rangle +
m^*\beta|00\rangle\right) \nonumber \\
& & \left. + |\Psi^{-}_{m}\rangle \left( mn\alpha |11\rangle -
\beta|00\rangle\right) \right].
\end{eqnarray}

\begin{figure}[!htb]
\centering
\includegraphics[width=6cm]{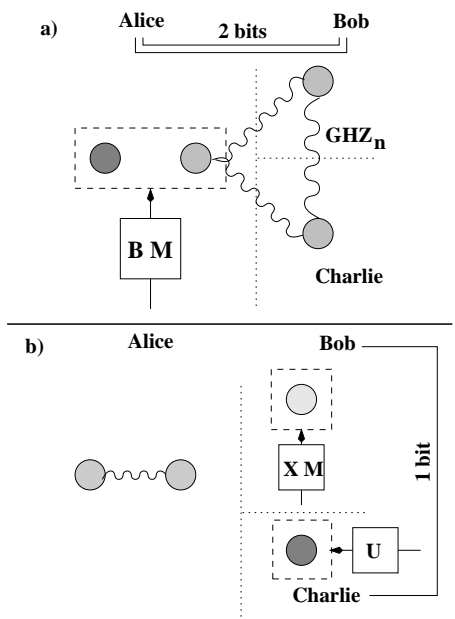}
\caption{a) Alice makes a generalized Bell measurement (BM) and informs Bob
and Charlie of the result ($2$ bits).
b) Bob makes an X-measurement (XM) and tells Charlie ($1$ bit), who applies a
proper unitary operation (U) on his qubit conditioned on Alice's and Bob's
measurement outcomes.}
\protect\label{Fig1-GHZ}
\end{figure}

Up to this point we have just rewritten Eq.~(\ref{initial-ghz}) in a convenient
form. The protocol begins when Alice implements a generalized Bell
measurement (BM) which is defined to be a projective measurement onto one of
the four  generalized Bell states, i.e Eqs.~(\ref{Bell1}-\ref{Bell2}). See
Fig.~\ref{Fig1-GHZ} for a pictorial representation of BM as well as of the
whole protocol.

For concreteness, let us assume Alice wants Charlie to receive the state.
(The protocol is symmetric up to this step and she can as well choose Bob to
receive the state.)
Therefore, after measuring her two qubits she tells Bob to implement on his
qubit an X-measurement (XM) which is a projective measurement on the
basis $|X^{\pm}\rangle=(1/\sqrt{2})(|0\rangle\pm |1\rangle)$. Alice and
Bob then tell Charlie their measurement outcomes. Depending on Alice's
and Bob's results Charlie implements a specific unitary operation, which are
all listed in Tab. \ref{table-ghz}, recovering Alice's state.
If Charlie does not learn from Bob
the outcome of the X-measurement his qubit will be left in a mixed state
for without that information Charlie cannot know the phase of Alice's qubit
\cite{Hil99}.
\begin{table}[!htb]
\caption{\label{table-ghz} The first column gives Alice's and Bob's outcomes,
the second one Charlie's unitary operation (UO), and the third Charlie's
qubit (unnormalized) at the end of the protocol. $I$ is the identity and
$\sigma$ Pauli matrices.}
\begin{ruledtabular}
\begin{tabular}{lcl}
BM and XM results & Charlie's UO & Charlie's qubit
\\ \hline
$\ket{\Phi^{+}_m}|X^+\rangle$ or $\ket{\Phi^{+}_m}|X^-\rangle$  &
$I$ or $\sigma_z$ & $\alpha|0\rangle + m^*n\beta|1\rangle$ \\
$\ket{\Phi^{-}_m}|X^+\rangle$ or $\ket{\Phi^{-}_m}|X^-\rangle$  &
$\sigma_{z}$ or $I$ &  $m\alpha|0\rangle + n\beta|1\rangle$ \\
$\ket{\Psi^{+}_m}|X^+\rangle$ or $\ket{\Psi^{+}_m}|X^-\rangle$ &
$\sigma_{x}$ or $\sigma_x\sigma_z$ &
$n\alpha|0\rangle + m^*\beta|1\rangle$ \\
$\ket{\Psi^{-}_m}|X^+\rangle$ or $\ket{\Psi^{-}_m}|X^-\rangle$ &
$\sigma_{z}\sigma_{x}$ or $\sigma_x$ & $mn\alpha|0\rangle + \beta|1\rangle$
\end{tabular}
\end{ruledtabular}
\end{table}

Looking at Tab. \ref{table-ghz} we see that Alice can achieve a unity fidelity
protocol by properly adjusting her measurement basis parameter $m$. For example,
if she chooses $m^*=1/n$ the protocol works when
her generalized BM gives $|\Phi^+_m\rangle$. There exist other three
possibilities: for $m=n$ the protocol succeeds when Alice's outcome is
$|\Phi^-_m\rangle$, for $m^*=n$ when she obtains $|\Psi^+_m\rangle$, and for
$m=1/n$ when she measures $|\Psi^-_m\rangle$. An interesting situation occurs
when $n$ is real \cite{footnote1}. Now, for $m=n$ the protocol works either
if Alice measures $|\Phi^-_m\rangle$ or $|\Psi^+_m\rangle$. Finally, for
$m=1/n$ the protocol works if Alice obtains either $|\Phi^+_m\rangle$ or
$|\Psi^-_m\rangle$.

\begin{figure}[!htb]
\centering
\includegraphics[width=6cm]{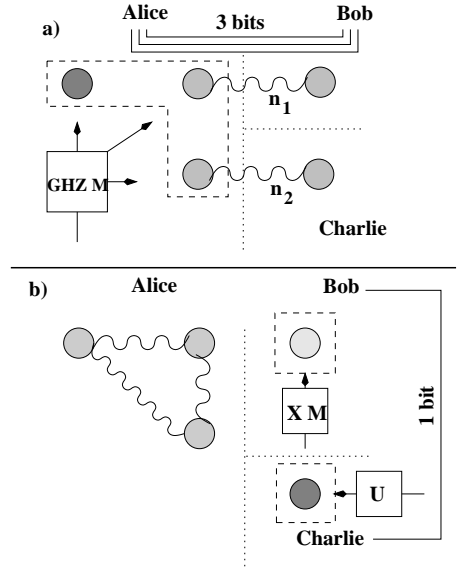}
\caption{a) Alice makes a generalized GHZ measurement (GHZ M) and informs Bob
and Charlie ($3$ bits) of the result.
b) Bob makes an X-measurement (XM) and tells Charlie ($1$ bit), who applies a
proper unitary operation (U) on his qubit conditioned on Alice's and Bob's
measurement outcomes.}
\protect\label{Fig2-Bell}
\end{figure}

We now turn our attention to the second QSTS protocol which is schematically
represented in Fig. \ref{Fig2-Bell}.

The main characteristic of the present protocol is its channels: two
non-maximally entangled Bell states. The initial state are composed of five
qubits. The first one, which is described by the state that will be
transferred to Bob or Charlie, the second,
and the fourth belong to Alice. The third and fifth
qubits are with Bob and Charlie respectively. The initial state then reads
\begin{equation}
|\Phi\rangle = | \phi \rangle_A \otimes |\Phi^+_{n_1}\rangle_{AB} \otimes
|\Phi^+_{n_2}\rangle_{AC},
\label{initial-bell}
\end{equation}
where subscripts were added to explicitly indicate which qubits are with Alice,
Bob, and Charlie. We now define the generalized GHZ basis as follows:
\begin{eqnarray}
|GHZ^{+}_{m}\rangle &=& M (|000\rangle + m |111\rangle),\label{GHZ1}\\
|GHZ^{-}_{m}\rangle &=& M (m^*|000\rangle - |111\rangle),\\
|G^{+}_{m}\rangle &=& M (|010\rangle + m |101\rangle),\\
|G^{-}_{m}\rangle &=& M (m^*|010\rangle - |101\rangle),\\
|H^{+}_{m}\rangle &=& M (|100\rangle + m |011\rangle),\\
|H^{-}_{m}\rangle &=& M (m^*|100\rangle - |011\rangle),\\
|Z^{+}_{m}\rangle &=& M (|110\rangle + m |001\rangle),\\
|Z^{-}_{m}\rangle &=& M (m^*|110\rangle - |001\rangle).
\label{GHZ2}
\end{eqnarray}
Defining $N_j=1/\sqrt{1+|n_j|^2}$ and using Eqs.~(\ref{GHZ1}) to (\ref{GHZ2})
we can rewrite the initial state (\ref{initial-bell}) as
\begin{eqnarray}
|\Phi\rangle & = & N_1N_2M \left[ |GHZ^+\rangle \left( \alpha|00\rangle +
m^*n_1n_2 \beta |11\rangle\right) \right.\nonumber \\
& & + |GHZ^-\rangle \left( m\alpha|00\rangle -n_1n_2 \beta |11\rangle\right)
\nonumber \\
& & + |Z^+\rangle \left( m^*n_2\alpha|01\rangle + n_1\beta |10\rangle\right)
\nonumber \\
& & + |Z^-\rangle \left( -n_2\alpha|01\rangle + mn_1\beta |10\rangle\right)
\nonumber \\
& & + |G^+\rangle \left( n_1\alpha|10\rangle + m^*n_2\beta |01\rangle\right)
\nonumber \\
& & + |G^-\rangle \left( mn_1\alpha|10\rangle - n_2\beta |01\rangle\right)
\nonumber \\
& & + |H^+\rangle \left( m^*n_1n_2\alpha|11\rangle + \beta |00\rangle\right)
\nonumber \\
& & \left. + |H^-\rangle \left( -n_1n_2\alpha|11\rangle
+ m\beta |00\rangle\right)\right].
\label{rewrite}
\end{eqnarray}
Note that in Eq.~(\ref{rewrite}) we have rearranged the qubits in order that
Alice's qubits are the first, the second, and the third ones and Bob's and
Charlie's are respectively the fourth and fifth ones.
The protocol begins by Alice implementing a generalized GHZ measurement (GHZ M)
in the sense that she projects her three qubits on one of the generalized GHZ
states.
Assuming she wants Charlie to receive the state if he collaborates with Bob,
Alice tells Bob to implement an X-measurement on his qubit. After that both
Alice and Bob inform Charlie of their measurement outcomes who ends the
protocol performing a unitary operation on his qubit depending on the
information he receives. For example, if Alice measures $|H^-\rangle$ and
Bob obtains $|X^+\rangle$ Charlie will need to implement $\sigma_x\sigma_z$ on
his qubit. Tab. \ref{table-bell} shows the final state with Charlie after he
has applied the proper unitary operation on his qubit. If Bob does not tell
Charlie his X-measurement result Charlie's qubit are left in a mixed state
since he has no information concerning the phase of Alice's qubit
\cite{Pen04}.
\begin{table}[!htb]
\caption{\label{table-bell} The first column gives Alice's and Bob's outcomes
and the second one Charlie's qubit (unnormalized) at the end of the protocol.}
\begin{ruledtabular}
\begin{tabular}{ll}
GHZ M and XM results & Charlie's qubit
\\ \hline
$\ket{GHZ^{+}_m}|X^{\pm}\rangle$ & $\alpha|0\rangle +
m^*n_1n_2\beta|1\rangle$ \\
$\ket{GHZ^{-}_m}|X^{\pm}\rangle$ & $m\alpha|0\rangle +
n_1n_2\beta|1\rangle$ \\
$\ket{G^{+}_m}|X^{\pm}\rangle$ & $n_1\alpha|0\rangle +
m^*n_2\beta|1\rangle$ \\
$\ket{G^{-}_m}|X^{\pm}\rangle$ & $mn_1\alpha|0\rangle +
n_2\beta|1\rangle$ \\
$\ket{H^{+}_m}|X^{\pm}\rangle$ & $m^*n_1n_2\alpha|0\rangle +
\beta|1\rangle$ \\
$\ket{H^{-}_m}|X^{\pm}\rangle$ & $n_1n_2\alpha|0\rangle +
m\beta|1\rangle$ \\
$\ket{Z^{+}_m}|X^{\pm}\rangle$ & $m^*n_2\alpha|0\rangle +
n_1\beta|1\rangle$ \\
$\ket{Z^{-}_m}|X^{\pm}\rangle$ & $n_2\alpha|0\rangle +
mn_1\beta|1\rangle$
\end{tabular}
\end{ruledtabular}
\end{table}

This time we have three parameters to play
with. In general we have no control over the channel entanglement. However,
the measuring basis parameter $m$ can be freely manipulated by Alice.
By proper adjusting it we can achieve a unity fidelity protocol. There
exist four possibilities. Looking at Tab. \ref{table-bell} we see
that for $m=n_1n_2$ the protocol works when Alice obtains either
$|GHZ^-_m\rangle$ or $|H^-_m\rangle$. However, when $m^*=1/(n_1n_2)$ we have
a successful run of the protocol if Alice measures either $|GHZ^+_m\rangle$
or $|H^+_m\rangle$. On the other hand, if $m^*=n_1/n_2$ a unity fidelity
transmission is achieved for $|Z^+_m\rangle$ or $|G^+_m\rangle$. Finally, for
$|Z^-_m\rangle$ and $|G^-_m\rangle$ the measurement basis parameter must
be set to $m=n_2/n_1$. It is worth mentioning that only if $n_1=n_2=m=1$ we
recover the protocol presented by Li \textit{et al.} \cite{Pen04}.

The security of both protocols against eavesdropping and cheating can be shown
by the same methods presented in Refs. \cite{Hil99,Pen04}. Actually, for the
successful instances of the protocols, i. e. those which Alice has correctly
adjusted her free parameter $m$, the same security tests developed for
the deterministic protocols \cite{Hil99,Pen04} apply. As a matter of fact,
it is Alice's ability to choose whether Bob or Charlie will receive the
transferred state which prevents cheating by one of the parties. If she thinks
one of the parties is the dishonest one, she can choose the other one to be the
receiver and by comparing a subset of the states received by the latter with
the states transmitted Alice can detect if the former is
cheating \cite{Hil99}.

We now compare the efficiency of both protocols employing the techniques
developed in Ref. \cite{Gor06}. We assume that any inefficiency of the
generalized BM and GHZ M is included in the following analysis by rescaling
the parameter $m$. Furthermore, from now on $n,m,n_1$, and $n_2$ are all real
numbers since it can be shown that we do not lose in generality by such
assumption \cite{Gor06}.
Each projective measurement implemented by Alice yields the state
$|R_j\rangle_A$ with probability $P_j$, where $|R_j\rangle_A$ stands for any
state Alice can measure. For each
one of Alice's and Bob's measurement outcomes and after implementing the proper
unitary operation Charlie ends up with the state $|\phi_j\rangle_C$. Therefore,
if $|\phi\rangle_A$ is the state Alice wanted to transfer the fidelity for
this run of the protocol is $F_j = |_A\langle \phi|\phi_j \rangle_C|^2$. In
general the probabilities $P_j$ and the fidelities $F_j$ depend on $\alpha$
and $\beta$. Moreover, Alice can change the values of $\alpha$ and $\beta$ of
the transferred state at will for each run of the protocol. Therefore, in order
to get $\alpha$- and $\beta$-independent results we average over many
implementations of the protocol obtaining the \textit{protocol efficiency}
\cite{Gor06}:
$
C^{pro} = \sum_j \langle P_j F_j\rangle.
$
In the averaging process we will need the quantities
$\langle |\alpha|^2 \rangle$,
$\langle |\alpha|^4 \rangle$, $\langle |\beta|^2 \rangle$,
$\langle |\beta|^4 \rangle$ and $\langle |\alpha\beta|^2 \rangle$. In Ref.
\cite{Gor06} they were shown to be $\langle |\alpha|^2 \rangle =
\langle |\beta|^2 \rangle = 1/2$,  $\langle |\alpha|^4 \rangle =
\langle |\beta|^4 \rangle = 1/3$, and $\langle |\alpha\beta|^2 \rangle = 1/6$.
We can interpret $C^{pro}$ as the average qubit transmission rate for a given
protocol choice \cite{Gor06}.

Assuming $|R_j\rangle_A=\{ |\Phi_m^+\rangle, |\Phi_m^-\rangle,
|\Psi_m^+\rangle,|\Psi_m^-\rangle \}$ for the first protocol we obtain
\begin{eqnarray}
C^{pro}_1 &=& \frac{2}{3}\left( 1 + \frac{2mn}{(1+m^2)(1+n^2)} \right)
\nonumber \\
&=&\frac{2}{3}\left( 1 + \frac{c(m)c(n)}{2} \right),
\label{C1}
\end{eqnarray}
where $c(m)=2|m|/(1+|m|^2)$ is the concurrence \cite{Woo98} of the generalized
Bell states \cite{Gor06}.
For the second protocol we have the following acceptable results
$|R_j\rangle_A=\{ |GHZ_m^{\pm}\rangle, |G_m^{\pm}\rangle,
|H_m^{\pm}\rangle, |Z_m^{\pm}\rangle \}$ and we get
\begin{eqnarray}
C^{pro}_2 &=& \frac{2}{3}\left( 1 + \frac{4mn_1n_2}{(1+m^2)(1+n_1^2)(1+n_2^2)}
\right) \nonumber \\
& = &  \frac{2}{3}\left( 1 + \frac{c(m)c(n_1)c(n_2)}{2}\right).
\label{C2}
\end{eqnarray}
If we compare Eqs.~(\ref{C1}) and (\ref{C2}) remembering that $0\leq
c(n)\leq 1$ we obtain
$
  C^{pro}_1 \geq C^{pro}_2
$
whenever $n=n_1$ or $n=n_2$. Therefore, for the same set of parameters the
first protocol is more efficient than the second one. We should mention that a
more complete efficiency analysis should also take account of the feasibility
of generating one GHZ state against two Bell states, which are the channels
of the first and second protocols respectively.

Eqs.~(\ref{C1}) and (\ref{C2}) also furnish other interesting
informations concerning each protocol. For example, for both
schemes we see that the protocol efficiency $C^{pro}$ is invariant
under the permutation of the parameters, $m$ and  $n$ for the
first QSTS protocol and  $m, n_1$, and $n_2$ for the second one.
In other words, if we interchange the degree of entanglement of a
channel ($n$, $n_1$, or $n_2$) with the measurement basis
entanglement degree ($m$) $C^{pro}$ is left unchanged. This same
result is also obtained for the generalized teleportation protocol
\cite{Gor06}. In a certain sense all these results suggest that
the entanglement of the channel and the entanglement of the
measuring basis are on equal footing in the determination of the
protocol efficiency. Moreover, both $C_1^{pro}$ and $C_2^{pro}$
increase either if we increase the degree of entanglement of the
channel ($n,n_1$, or $n_2$) or the measuring basis entanglement
($m$), an expected result since by increasing the quantum resource
(entanglement) available we should improve the efficiency of the
protocols. And only when $m=n=1$ or $m=n_1=n_2=1$ we achieve unity
efficiency and recover the protocols presented by Hillery
\textit{et al.} \cite{Hil99} and Li \textit{et al.} \cite{Pen04}.
Furthermore, the dependency of the efficiency on the entanglement
resource enables one to compare the two protocols and the channels
used. This can be quantified by evaluating $c(n)$ and
$c(n_1)c(n_2)$ which may lead to a new way of comparing
multipartite entanglement with pairwise entanglement.

Alice can easily extend (at least theoretically) the previous protocols to
transfer her qubit to a specific party among $N-1$ parties. For the first
protocol she needs to share with the parties a $N$-qubit GHZ state,
$|GHZ_N\rangle = (1/\sqrt{2})(|0\rangle^{\otimes N} + |1\rangle^{\otimes N})$,
as the channel. She then implements a generalized BM and asks all the
parties but the one chosen to receive the state to make an X-measurement on
their qubits. Then,  if the chosen party receives the results of the $N-2$
X-measurements and Alice's outcome he can recover the state by applying
proper unitary operations on his qubit.
For the second protocol, Alice needs to share $N-1$ Bell states, each one
with each party. Then she implements a generalized GHZ M on her $N$ qubits:
the one to be transferred and the $N-1$ from the Bell states. The rest of
the protocol works as the previous one: all but the chosen party make
X-measurements on their qubits and the receiver obtains the transferred state
by applying a unitary operation on his qubit conditioned on the information
received from Alice and the other $N-2$ parties.

We end this contribution noting that in general decoherence and noise degrade
the entanglement of the channel in a rather complicated way. Most of the time
an initially pure state (or equivalently pure channel) evolves non-unitary to
a mixed state. Here, however, we restricted ourselves to a ``unitary loss''
of entanglement, in which a maximally entangled pure channel evolves to a
partially entangled one:
$$
\frac{1}{\sqrt{2}} (|00\rangle + |11\rangle) \longrightarrow
\frac{1}{\sqrt{1+|n|^2}}(|00\rangle + n|11\rangle).
$$
Note that the bit flip noise $|0\rangle \longrightarrow |1\rangle$ is also a
unitary noise, although it does not change the entanglement of the channel.
Charlie can easily overcome it, and obtain the states shown in Tabs.
\ref{table-ghz} and \ref{table-bell}, by implementing a proper unitary
operation on his qubit at the end of the protocol.

In order to attack a non-unitary loss of entanglement, where a
pure channel evolves to a mixed channel, a more subtle and
sophisticated approach is needed. The present treatment will be
extended elsewhere to include this more realistic scenario in a
concise and general way.

The authors thank G. Kurizki for his support and guidance.
G. R. thanks FAPESP for partially funding this research.


\begin{thebibliography}{99}

\bibitem{Hil99} M. Hillery, V. Bu\v{z}ek, and A. Berthiaume, Phys. Rev. A
\textbf{59}, 1829 (1999).
\bibitem{Lo99} R. Cleve, D. Gottesman, and H.-K. Lo, Phys. Rev. Lett.
\textbf{83}, 648 (1999).

\bibitem{Pen04} Y. Li, K. Zhang, and K. Peng, Phys. Lett. A \textbf{324},
420 (2004).
\bibitem{Lan04} A.M. Lance, T. Symul, W.P. Bowen, B.C. Sanders, and P.K. Lam,
Phys. Rev. Lett. \textbf{92}, 177903 (2004).

\bibitem{Den05a} F.-G. Deng, C.-Y. Li, Y.-S. Li, H.-Y. Zhou, and Y. Wang,
Phys. Rev. A \textbf{72}, 022338 (2005).
\bibitem{Den05b} F.-G. Deng, X.-H. Li, C.-Y. Li, P. Zhou, and H.-Y. Zhou,
Phys. Rev. A \textbf{72}, 044301 (2005).
\bibitem{Kar99} A. Karlsson, M. Koashi, and N. Imoto, Phys. Rev. A \textbf{59},
162 (1999).

\bibitem{Gis01} W. Tittel, H. Zbinden, and N. Gisin, Phys. Rev. A \textbf{63},
042301 (2001).

\bibitem{Cir97} J.I. Cirac, P. Zoller, H.J. Kimble, and H. Mabuchi, Phys. Rev.
Lett. \textbf{78}, 3221 (1997).

\bibitem{Ben96} C.H. Bennett, G. Brassard, S. Popescu, B. Schumacher,
J. A. Smolin, and W.K. Wootters, Phys. Rev. Lett. \textbf{76}, 722 (1996).

\bibitem{Agr02} P. Agrawal and A.K. Pati, Phys. Lett. A \textbf{305},
12 (2002).

\bibitem{Gor06} G. Gordon and G. Rigolin, Phys. Rev. A \textbf{73}, 042309
(2006).

\bibitem{footnote1} Even if the channel parameter $n$ is complex we can
cancel its phase obtaining a real parameter by allowing Charlie to implement
another unitary operation \cite{Gor06}.

\bibitem{Woo98} W. K. Wootters, Phys. Rev. Lett. \textbf{80}, 2245 (1998).


\end{thebibliography}
\end{document}